**Understanding alkali contamination in colloidal nanomaterials to unlock grain boundary impurity engineering**

Se-Ho Kim[§,†,*], Su-Hyun Yoo[§,†,*], Poulami Chakraborty[§], Jiwon Jeong[§], Joohyun Lim[§,‖], Ayman A. El-Zoka[§], Xuyang Zhou[§], Leigh T. Stephenson[§], Tilmann Hickel[§], Jörg Neugebauer[§], Christina Scheu[§], Mira Todorova[§], Baptiste Gault[§,‡,*]

[§]Max-Planck-Institut für Eisenforschung GmbH, Max-Planck-Straße 1, 40237 Düsseldorf, Germany

[‖]Department of Chemistry, Kangwon National University, Chuncheon 24342, Republic of Korea

[‡]Department of Materials, Royal School of Mines, Imperial College, London, SW7 2AZ, United Kingdom

[†]these authors contributed equally

[*]corr. Authors: s.kim@mpie.de | yoo@mpie.de | b.gault@mpie.de

## Abstract

Metal nano-gels combine a large surface area, a high structural stability, and a high catalytic activity towards a variety of chemical reactions. Their performance is underpinned by the atomic-level distribution of their constituents, yet analysing their sub-nanoscale structure and composition to guide property optimization remains extremely challenging. Here, we synthesized Pd nano-gels using a conventional wet chemistry route, and near-atomic-scale analysis reveals that impurities from the reactants (Na, K) are integrated into the grain boundaries of the poly-crystalline gel, typically loci of high catalytic activity. We demonstrate that the level of impurities is controlled by the reaction condition. Based on *ab initio* calculations, we provide a detailed mechanism to explain how surface-bound impurities get trapped at grain boundaries forming as the particles coalesce during synthesis, possibly facilitating their decohesion. If controlled, impurity integration into grain boundaries may offer opportunities for designing new nano-gels.

## Introduction

Despite over 150 years 'wet-' chemical synthesis of colloidal metal nanoparticles and other nanostructures remains, many aspects of their structure and composition remain elusive, leading to Xia et al. saying that it remains 'an art rather than a science'[1]. The wet-synthesis of nanostructures typically involves adding a reducing agent to a solution containing a metal precursor. Due to its excellent reducing properties[2], $NaBH_4$ is the most commonly used in laboratory and industrial applications[3] – the so-called $NaBH_4$ reduction method[4]. Early nuclei of nano-crystals form and grown by agglomeration of the reduced metal atoms, becoming stable nano-crystals upon reaching a critical size[5]. Yet alkalis' (Na or K) interaction with the growing nanocrystals is rarely considered in growth models.

The rapid generation of metal atoms following the introduction of a strong reducing agent results in a high concentration of crystal nuclei that eventually coalesce into a nano-gel and become a metal nano-aerogel structure (MNA)[6,7]. MNAs are an emerging class of self-supported porous materials with potential in electrocatalysis, surpassing commercial metal-based catalysts because of their structural stability and efficient mass/electron transfer channels[8,9]. MNAs are extensively studied across an array of catalytic applications such as oxygen reduction reaction[10,11], glucose oxidation reaction[12,13], and ethanol oxidation reaction[14,15].

MNAs synthesized by $NaBH_4$ reduction of a metal precursor are often perceived as purely metallic, i.e. without impurities integrated in their complex nano-porous structure[16,17]. Impurities inside MNAs can modify their stability, electronic structure, and hence their reactivity. Here, we study the controlled integration of impurities into nano-gels by combining microscopy and microanalysis at the near-atomic scale[18,19], with *ab initio*-density functional theory (DFT) calculations to investigate the energetics and incorporation of impurities into the nanostructures, in particular at crystalline defects. The mechanism we outline will facilitate tailoring MNAs for specific catalytic reactions by using insights from atomistic simulations [20].

## Results and Discussion

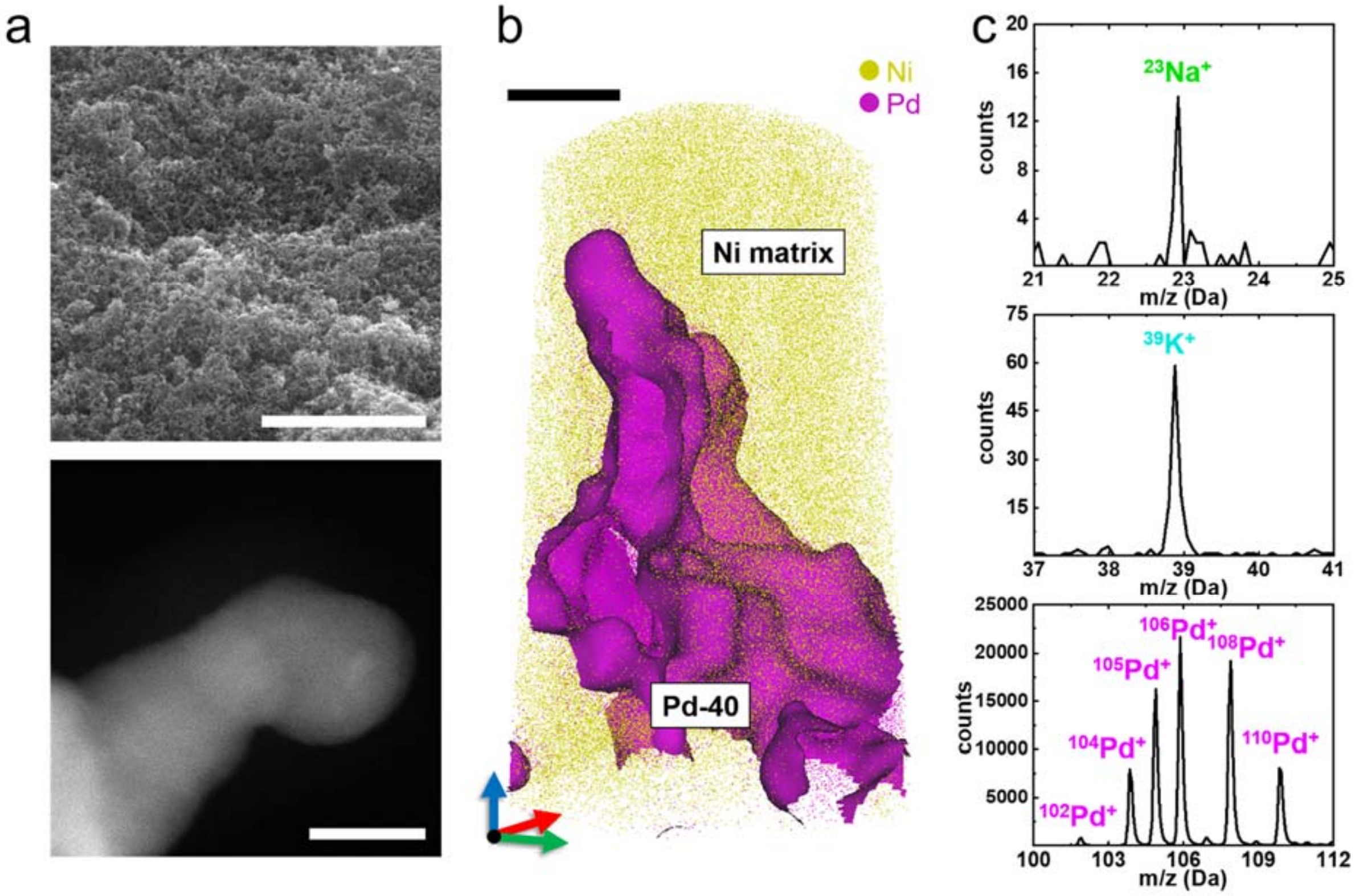

**Figure 1.** Characterization of Pd-40 nano-gels. (a) SEM and HAADF-STEM images of as-synthesized Pd-40 gels. Scale bars are 5 μm and 10 nm for the SEM and STEM image, respectively. (b) 3D atom map of Pd-40 gels embedded in a Ni matrix (scale bar = 10 nm). A one-dimensional composition profile positioned perpendicular to the matrix-MNA interface is shown in Figure S3. (c) Illustration of major and minor peaks in three different mass-to-charge ratio ranges: Na, K, and Pd. Overall mass spectrum is presented in Figure S4.

We synthesized two Pd-gels with two mole ratios of $NaBH_4$ reductant to the Pd-precursor (R/P): 40 and 0.1, referred to as Pd-40 and Pd-0.1, respectively, (for details, please see Experimental Section). After synthesis, the gels were thoroughly washed three times with distilled water to remove surface residuals. Figure 1a displays micrographs of the as-synthesized Pd-40 obtained by scanning electron microscopy (SEM) and high-angle annular dark field-scanning transmission electron microscopy (HAADF-STEM). The three-dimensional, highly complex network of pores of the MNA structures is readily apparent, with an average ligament size of approx. 15 nm. Energy-dispersive X-ray spectroscopy (EDS) (Figure S1 and S2) hints to the presence of Na and K. The complex geometry of the specimen and low concentration of the impurity elements made extremely challenging the combined quantification and highly spatially-resolved localization within the MNA[21,22].

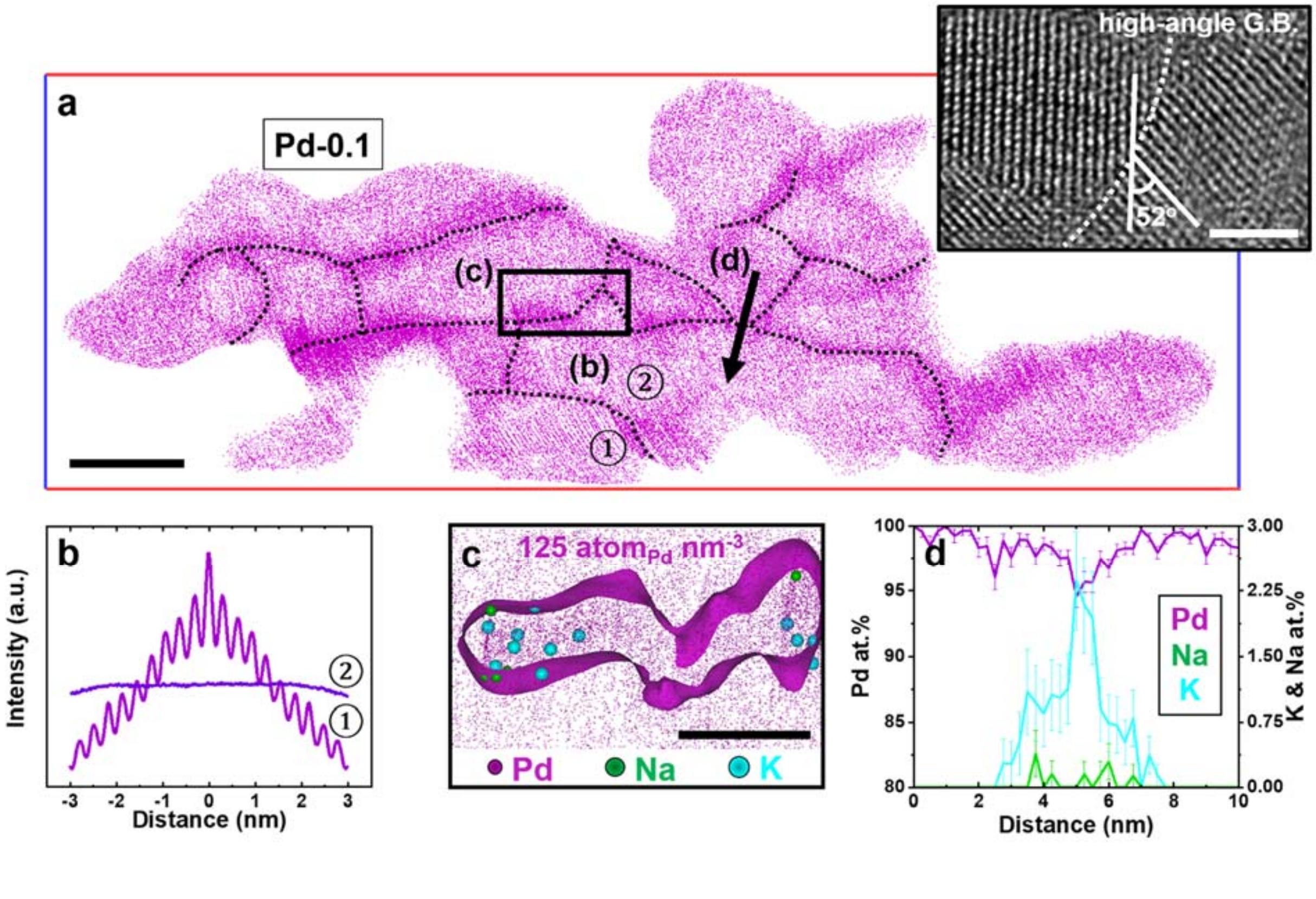

**Figure 2.** Grain boundary studies for Pd-0.1 nano-gels. (a) A one-nanometer thin sliced tomogram from a 3D atom map (Figure S5) of Pd-0.1 gels (iso-composition surface >90 at.% Pd). The scale bar is 10 nm. The dotted black line represents grain boundary features. Inset shows a high-resolution TEM image of Pd-0.1 (scale bar = 2 nm). (b) Spatial distribution maps along the z-axis from two different nano-gel's grains. (c) Extracted grain boundary tomogram with the iso-density surface of 125 Pd atoms per nm$^3$. The scale bar is 5 nm. (d) 1D compositional profiles of detected Na, K, and Pd elements.

We then performed atom probe tomography (APT) using the protocol outlined in ref.[23] and embedded the MNA in a Ni matrix (see Experimental Section). The 3D atom map for the Pd-40 embedded in Ni is shown in Figure 1b. An iso-composition surface of 50 at.% Pd highlights a complex morphology of Pd-40 compatible with the STEM imaging. The APT mass spectrum from the Pd-rich region, Figure 1c, contains peaks associated with Na and K at 23 and 39 Da, respectively, along with the isotope of singly charged Pd (102 to 110 Da). These impurities, Na and K, likely originate from the reducing agent ($NaBH_4$) and the Pd-precursor ($K_2PdCl_4$). Although alkali metals are often believed to be surface residuals on nanoparticle system, the Na and K are detected inside the Pd-40 network with a composition of 94±2 and 512±50 atomic parts per million (appm), respectively. Cl only weakly binds to Pd-surfaces in aqueous solution[24], and no Cl is detected.

To unveil the origin of these impurities, we studied the Pd-0.1 (*i.e.* R/P = 0.1), *i.e.* the same concentration of Pd-precursor but less $NaBH_4$ in the solution. The morphology and ligament size are similar to Pd-40 (Figure S5 and S6). Du *et al.*[6] reported an influence of the R/P ratio on the ligament size of Au MNAs, with a low (<2) and high (>50) R/P. Here, the Pd-precursor solution is more highly concentrated (0.01 M *vs.* 0.0002 M), leading to a higher concentration of nuclei to form the MNA. Pd-0.1 contains a similar amount of K (512 *vs.* 642 appm) in both MNAs, and their synthesis involved the same concentration of the $K_2PdCl_4$ precursor. In contrast, the lower concentration of Na-containing reductant in solution leads to a lower incorporation of Na (22 *vs.* 94 appm). Table S1 reports the compositions of both samples.

A new batch with one-to-one mole ratio, Pd-1 was also synthesized. Figure S7 shows the 3D atom maps and sectioned tomogram of the Pd-1 sample. Inside of the gel structure, Na and K are detected at 73±7 and 417±16 appm levels, respectively. A similar amount of K is detected as the same concentration of the precursor was involved whereas the Na content is between that of Pd-0.1 and Pd-40 samples (see Figure S8).

During the MNA synthesis, following the initial growth of nuclei reaching a critical size, the crystals start to coalesce[25] and merge with each other with no particular crystallographic relationship, conversely to orientated attachment[26]. Therefore, MNAs are agglomerations of randomly oriented crystal grains with numerous interfaces, *i.e.* grain boundaries (GBs)[27]. A one-nanometer thick slice through the APT reconstruction, Figure 2a, reveals the presence of such GBs in the Pd MNA. The apparent higher point density at the junction nanocrystals is related to aberrations in the ion trajectories[28]. GBs are marked by black dotted lines. In Figure S9 and the inset in Figure 2a, high resolution (HR-) TEM image shows multiple high-angle grain boundaries (>15°) within the poly-crystalline nature of Pd nano-gels. In addition, grain ① shows a periodic arrangement of atoms pertaining to a set of crystallographic planes in the APT data, as revealed by the so-called spatial distribution map[29] plotted in Figure 2b, which terminates at the junction with the neighboring grain ②, *i.e.* each grain in Pd-0.1 has a different orientation. Na and K are mostly found along the GBs, Figure 2c, and a composition profile calculated along the black arrow, Figure 2d, indicates that the GB contains 0.5 at.% Na and 2.3 at.% K. The segregation to interfaces is often reported in bulk materials[30,31] but had not been studied in MNAs.

To rationalize the impurity incorporation into the MNAs, we first evaluate the binding energy ($E_b$) of Na and K adsorbates relative to their respective BCC bulk phase (zero reference for the chemical potential), at different binding sites on a Pd (111) surface using density functional theory (DFT). The modelled Pd surface is a proxy for a crystal nucleus, *i.e.* before coalescence. The 3-fold hollow sites (i.e., FCC and HCP) are the most favorable binding sites for both alkalis at low coverages ($\Theta \leq 0.25$), Figure 3a and 3b, and for both, the surface binding energy monotonously reduces with increasing coverage. We used the Nernst equation to calculate the chemical potentials from the ion concentration in solution for two synthesis conditions, and they are shown as colored horizontal dashed lines in each pane of Figure 3. The difference between $E_b$ and the corresponding chemical potential for low and high concentration in solution (blue- and red-dashed line, resp.) is the net binding energy, which allows us to directly determine the equilibrium concentration of the alkali atoms at the surface. Figure 3c indicates that the surface concentration of both Na and K is large, at 0.31 ML for K, and 0.74 to 0.52 ML for Na, when reducing the Na concentration in solution from 0.4 to 0.001 M.

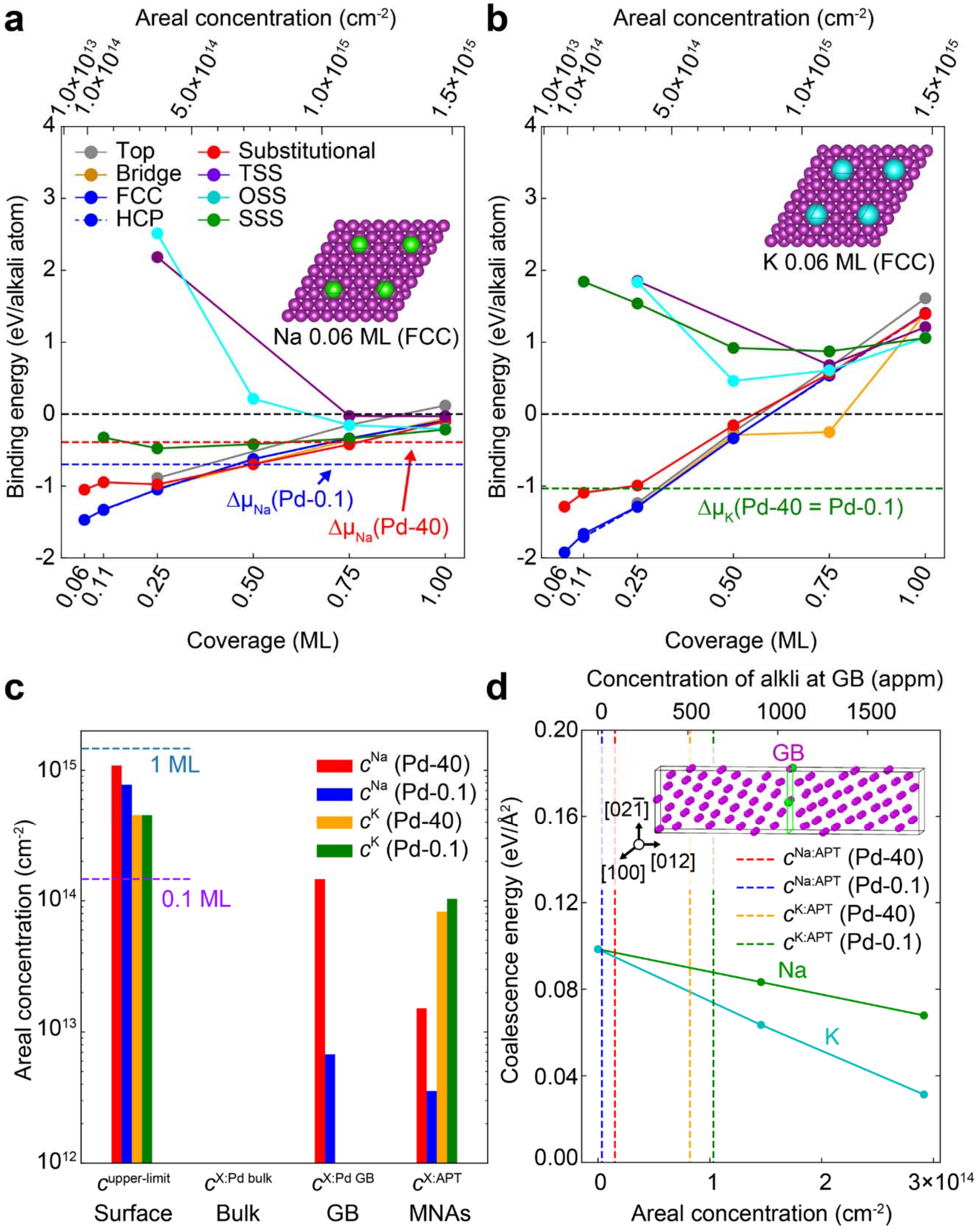

**Figure 3.** Surface adsorption and grain boundary energy calculations. The binding energies with respect to the alkali BCC bulk phase ($E_b$) of (a) Na and (b) K adsorbates at the Pd (111) surface and sub-surface for several adsorbate coverages in the range 0.06 ML to 1 ML. Solid lines of different colors correspond to different binding sites for the alkali atom. Colored horizontal dashed lines show the shifts in the chemical potentials corresponding to alkali ions in solution for the given experimental conditions, with respect to the alkali BCC bulk phase. Insets show examples of binding at FCC sites for 0.06 ML of Na (green) and K (cyan) on the *p*(2x2) Pd (purple) surface. (c) Plotted in a log scale are the thermodynamic equilibrium concentrations of alkalis on the surface at 300 K, in the bulk and in the GB, and the experimentally measured concentrations of alkalis in MNAs for the considered experimental conditions. Horizontal dashed lines indicating areal concentrations of 1 and 0.1 ML surface coverages are shown as guides. (d) The coalescence energy of the GB is plotted as a function of areal concentration of alkalis in the GB (assuming the concentration of alkalis in the GB to be identical to that on the surface), with the experimentally observed alkali contents shown as vertical dashed lines. The inset shows the supercell containing the Σ5 (012) Pd grain boundary used in the coalescence energy calculations.

APT analysis reveals Na and K atoms predominantly at GBs, not at the surface, from which they were likely leached by rinsing with water. To rationalize the presence of alkalis at GBs within Pd-MNAs, we first consider two extreme scenarios: (i) all alkali atoms chemisorbed at the surface become trapped during coalescence, and, (ii) the alkalis at the GBs can achieve a thermodynamic equilibrium with those in solution. Since coalescence brings two surfaces together the GB concentration for case (i) would be twice the surface concentration. Since the experimentally observed Na and K concentration at GBs are significantly lower, at least a (partial) equilibration takes place.

With regards to the second scenario, we computed by DFT the GB concentration of Na and K, in equilibrium with the corresponding chemical potential in solution, for a Σ5 (210) [001] GB, with a misorientation of 53.13˚ and an open structure (Figure 3c and supporting information). Assuming that this Σ5 GB can serve as a proxy for a random high-angle GB, the computed equilibrium GB concentration of Na agrees qualitatively with experimental concentrations. For K, however, the measured concentration is orders of magnitude than predicted ($1.0\times10^{14}$ $cm^{-2}$ *vs* $4.2\times10^{-12}$ $cm^{-2}$). The high concentration of K atoms initially present at the surfaces is only partially released back in solution to achieve thermodynamic equilibrium during coalescence, and that a substantial number of K atoms are kinetically trapped in the GB plane (roughly 20% for Pd-40 and 2% for Pd-0.1). The deduced sluggish kinetics of K compared to Na is also supported by a recent report for polycrystalline Mo and Nb that the larger-sized K has a smaller diffusion coefficient than Na by a factor of two to three[32].

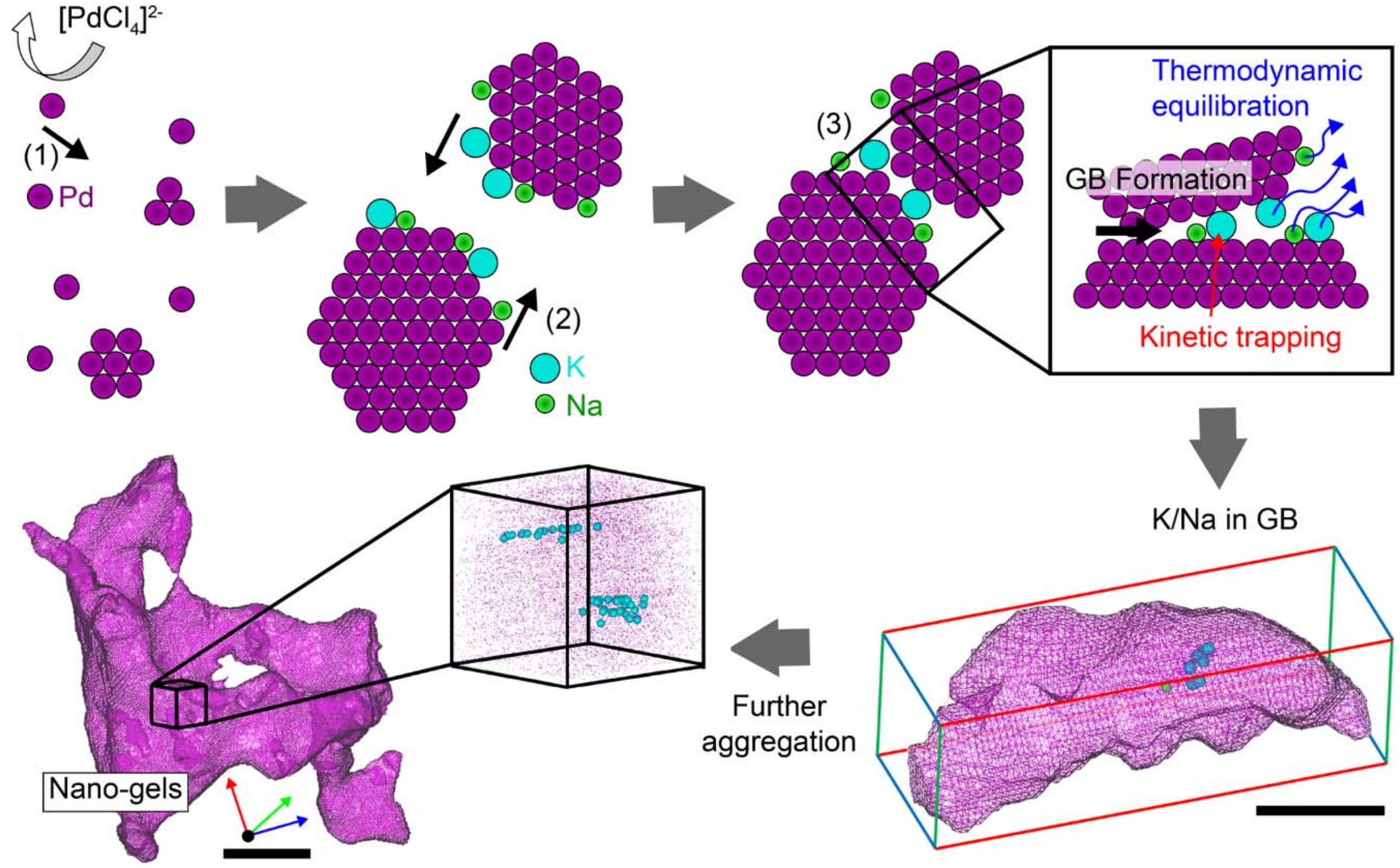

**Figure 4.** Alkali atoms integration mechanism. Schematic illustration of steps comprising the formation mechanism of Pd nano-gels: reduction of Pd atoms, formation of a Pd nuclei, coalescence of Pd nano-crystals, aggregation of primary Pd nano-gels, and final Pd nano-gels. The last two steps are from APT results (scale bars are 5 and 10 nm for the 4$^{th}$ and 5$^{th}$ step, respectively). Purple wireframes represent iso-composition surface of Pd at >90 at.%. Green and cyan dots represent Na, and K atoms, respectively. During the nano-gel formation, (1) reduction of the Pd precursor to Pd$^{o}$ atom (purple), (2) absorption of alkali atoms [K (cyan) and Na (green)], and (3) coalescence of Pd nano-crystals with integration of alkali atoms into the interface occur rapidly.

We then studied the GB coalescence energy, i.e. the energy required to form a GB interface from two surfaces, as function of Na and K concentration. Indeed, a large concentration of alkali atoms could energetically stabilize these metallic surfaces so much that forming a GB would be thermodynamically unfavorable and coalescence would be suppressed. As shown in Figure 3d, the coalescence energy is systematically positive, coalescence is hence energetically favorable. K almost halves the GB coalescence energy, whereas Na has relatively less influence.

From the perspective of the mechanical properties, the embrittlement of grain boundaries can result in the catastrophic facture of crystalline materials[30] and would likely affect MNAs' lifetime and device durability. In contrast, since the presence of K changes the energetics of GBs and hence their likelihood of formation, using an excess amount of K during synthesis could lead to an increase in the surface area-to-volume ratio, and hence be beneficial for catalysis applications. We demonstrate this qualitatively in Figure S10, in which STEM and APT analysis of a Pd-MNA synthesized with an excess of K from KCl, shows relatively thinner ligaments.

## Conclusions

To conclude, our theoretical and experimental insights allow us to propose a general mechanism for alkali integration in nanostructures synthetized by wet chemistry, schematically illustrated in Figure 4. Nanocrystals form throughout the solution, with their metal-solution interface partly stabilized by impurities, *e.g.* K and Na. Through agglomeration of these nanocrystals, a large number of impurities end up at the generated internal interfaces, *i.e.* grain boundaries. Impurities are then leached back in solution to reduce the gel's internal energy. During this process, interfaces become increasingly more stable, but a substantial amount of the large-sized K atoms remains kinetically trapped by the moving surfaces of the growing nanocrystals.

Importantly, this opens the possibility of using grain boundary engineering (GBE) guided by information obtained in silico for property optimization of colloidal nanomaterials. In oxide poly-crystals, the influence of space charge due to metal excess at the interfaces is well studied. The dangling bonds of the excess metal in oxides will lead to bond electrons at the interface and fill the electronic states near the conduction band which lowers the material intrinsic band-gap[33–35]. GBE successfully allowed control over the mechanical and chemical properties of bulk metallic alloys (e.g. steel[36] and Mg alloy[31]) and inorganic compounds (e.g. CIGS[37]).

Our results show the potential in applying it to decorate GBs in freestanding nanomaterials. For example, the local change in the electronic properties at GBs in MNAs[38] facilitates the adsorption of CO2 and provide fast kinetics for CO2 reduction reaction[39,40]. Properties could also be further enhanced by promoting the adsorption of specific elements selected from *ab initio* calculations. We also showed that the concentration of these impurities within the nanostructure is controlled by their initial concentration in solution and the relative energetics of the surface and GBs, providing levers to help future material design.

Eliminating the source of alkali (*e.g.* $NaBH_4$ reductant and/or potassium-based metal precursor) in MNA synthesis adversely affects the high production rate, and, therefore, there is an inevitable inverse relation between impurities integration and efficient MNA production. In addition, impurities at GB could reduce its cohesion, which is detrimental to the longevity of MNAs. Property optimization will hence depend on a subtle compromise, and the effective usage and/or removal of alkali atoms at GBs during coalescence appears to be pivotal for a successful application of GBE in MNA. There may also be opportunities to exploit impurity ingress to dope the material in order to obtain a strengthening effect to counteract the coalescence, or as a promotor to the catalytic activity.

**Experimental Section**

**Synthesizing Pd nano-gels.** 0.01 M $K_2PdCl_4$ (potassium tetrachloropalladate 99.99%, Sigma Aldrich) was mixed with 0.4 M of $NaBH_4$ (sodium borohydride, 99.99%, Sigma-Aldrich) for Pd-40 synthesis (for Pd-0.1, 0.001 M of $NaBH_4$ is used). After the reaction completely stopped, a centrifuge was used to collect the Pd black powder, which was re-dispersed in distilled water. This process was carried out three times to remove excess residuals on the nano-gels. Finally, the collected powder was then dried in a vacuum desiccator for a day.

**Sample preparation for APT measurement.** As-synthesized Pd nano-gels were prepared into a APT sample following the modified co-electrodeposition technique[41,42]. For an electrolyte preparation, Nickel (II) sulfate hexahydrate (98%, Sigma-Aldrich) and citric acid (99.5%, Sigma-Aldrich) were dissolved in 50 mL of distilled water. Here we used H-citric acid to avoid any possible Na or K introduction from the APT sample process. A constant current of -38 mA for 1250 sec was applied to completely encapsulate colloidal Pd nanomaterials with Ni film.

**TEM characterization.** HAADF-STEM images were acquired using a JEM-2200FS TEM (JEOL) at 200 kV. Elemental mapping using EDS was carried out for the investigation of the chemical composition of Pd nano-gels. HRTEM images were obtained with an aberration-corrected FEI Titan Themis 60-300 microscope at 300 kV.

**APT characterization.** Pd gels/Ni APT specimens were prepared using the standard specimen preparation technique[43] with focused ion beam (FEI 600 DualBeam) and subsequently were loaded inside a LEAP 5000 XS (CAMECA). APT measurement was performed in pulsed laser mode at set temperature of 50 K. The detection rate of 1 %, laser pulse frequency of 200 kHz, and laser energy of 60 pJ were used throughout the measurement. Acquired data set were then analyzed using IVAS 3.8.4 (CAMACA) software.

**Computational Calculation**. The Vienna *Ab initio* Simulations Package (VASP) code[44] employing the projector augmented wave (PAW) method[45] is used for all DFT calculations. A plane-wave cutoff of 500 eV is used, which is sufficient to achieve force convergence of 0.01 eV/Å and total energy convergence of $10^{-6}$ eV. The generalized gradient approximation (GGA) due to Perdew, Burke, and Ernzerhof[46] is used for the exchange-correlation approximation. Brillouin-zone integration is carried out using Methfessel–Paxton smearing. Γ-centered *k*-point grids with the following *k*-points are used for Brillouin-zone integrations: (8×8×8) for face-centered cubic Pd bulk, (8×8×1) for the Pd(111) *p*(1×1) surface unit cell and (2×9×9) for grain boundary calculations. The *k*-point meshes in the surface calculations are equivalently folded according to the size of the considered surface cells. Electronic and ionic relaxations are carried out until the total energy convergence is less than $10^{-5}$ eV, respectively $10^{-4}$ eV per system. Within this setup, the obtained lattice parameter $a$ = 3.959 Å and cohesive energy $E_{coh}$ = 3.63 eV of Pd fcc bulk agree well with previous theoretical[47,48] and experimental[49] results. Details on surface adsorptions and grain boundaries calculations and concentration analysis are presented in the SI.

## Author Contributions

S.-H.K. and B.G. designed the study. S.-H.K. performed the synthesis, co-electrodeposition, and atom probe specimen preparation. S.-H.K. measured atom probe tomography and analyzed the acquired data with support from A.A.E., L.T.S., and B.G.. J.J. and J.L. performed HR-TEM and (S)TEM-EDS measurements with support from C.S.. X.Z. analyzed the EELS measurement. S.-H.Y. and P.C. performed DFT calculations with support from T.H., M.T., and J.N.. S.-H.K., S.-

H.Y. and B.G. drafted the manuscript. All authors then contributed and have given approval to the final version of the manuscript.

## Notes

The authors declare no competing financial interest.

## Acknowledgement

We thank Uwe Tezins, Christian Broß, and Andreas Sturm for their support to the FIB and APT facilities at MPIE. S.-H.K. and B.G. acknowledge financial support from the ERC-CoG-SHINE-771602.

**Supporting Information**

**Understanding alkali contamination in colloidal nanomaterials to unlock grain boundary impurity engineering**

Se-Ho Kim[§,†,*], Su-Hyun Yoo[§,†,*], Poulami Chakraborty[§], Jiwon Jeong[§], Joohyun Lim[§,‖], Ayman A. El-Zoka[§], Xuyang Zhou[§], Leigh T. Stephenson[§], Tilmann Hickel[§], Jörg Neugebauer[§], Christina Scheu[§], Mira Todorova[§], Baptiste Gault[§,‡,*]

[§]Max-Planck-Institut für Eisenforschung GmbH, Max-Planck-Straße 1, 40237 Düsseldorf, Germany

[‖]Department of Chemistry, Kangwon National University, Chuncheon 24342, Republic of Korea

[‡]Department of Materials, Royal School of Mines, Imperial College, London, SW7 2AZ, United Kingdom

[†]these authors contributed equally

[*]corr. Authors: s.kim@mpie.de | yoo@mpie.de | b.gault@mpie.de

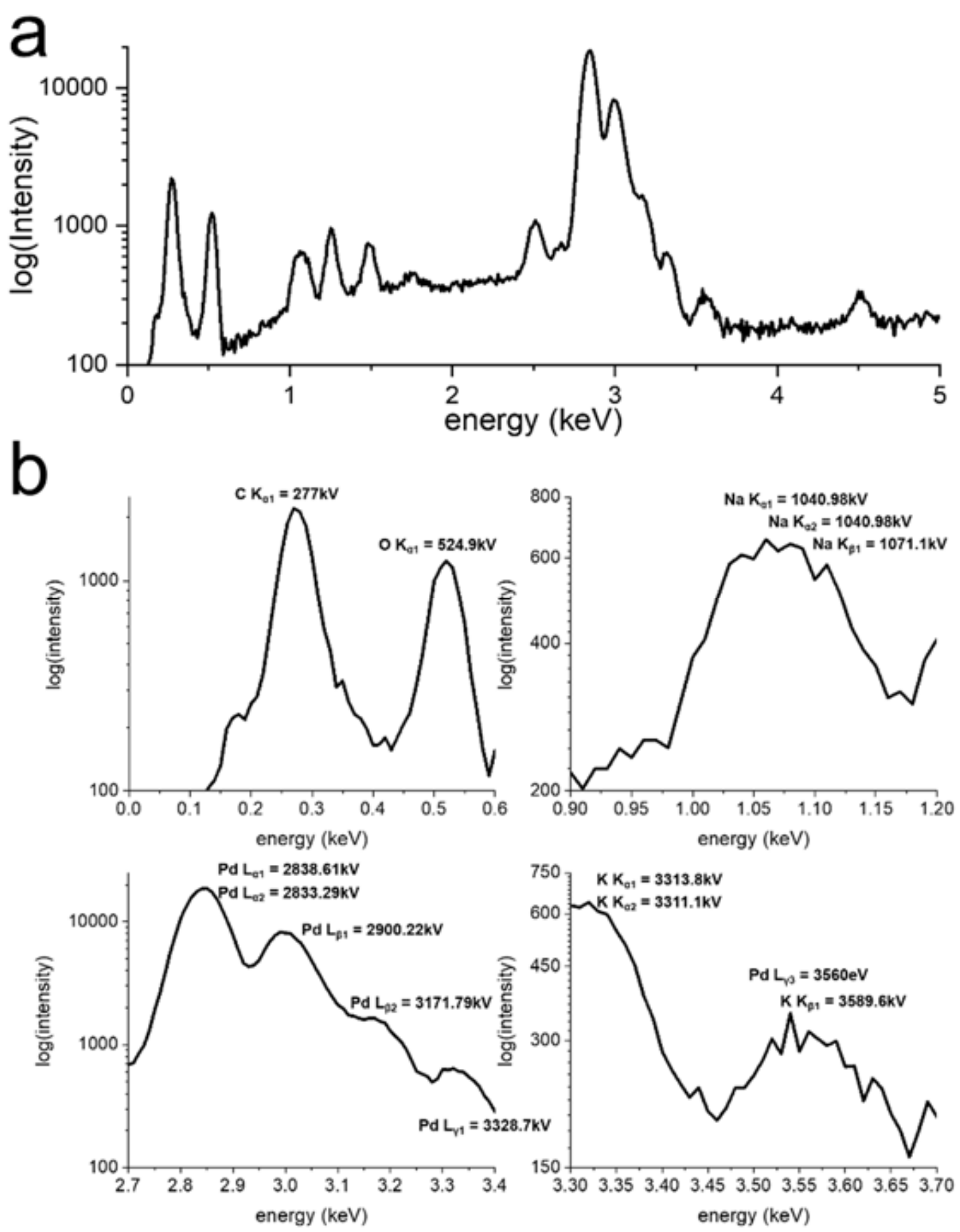

**Figure. S1.** (a) Overall energy-dispersive X-ray spectroscopy (EDS) spectrum of the as-synthesized Pd-40 acquired from scanning electron microscopy and (b) illustration of major peaks in four different ranges. Each characteristic X-ray peak signal is interpreted following ref.[1]. Along with strong signals of Pd $L_\alpha$, low intensities of the characteristic X-ray peak for Na $K_\alpha$, and K $K_\beta$ are detected; however, due to the high background levels and peak signal overlaps, for instance between K $K_\beta$ (3.57 keV) and Pd $L_{\gamma 3}$ (3.56 keV)[1,2], acquiring good chemical resolution for the Pd nano-gels was difficult from X-ray based characterization.

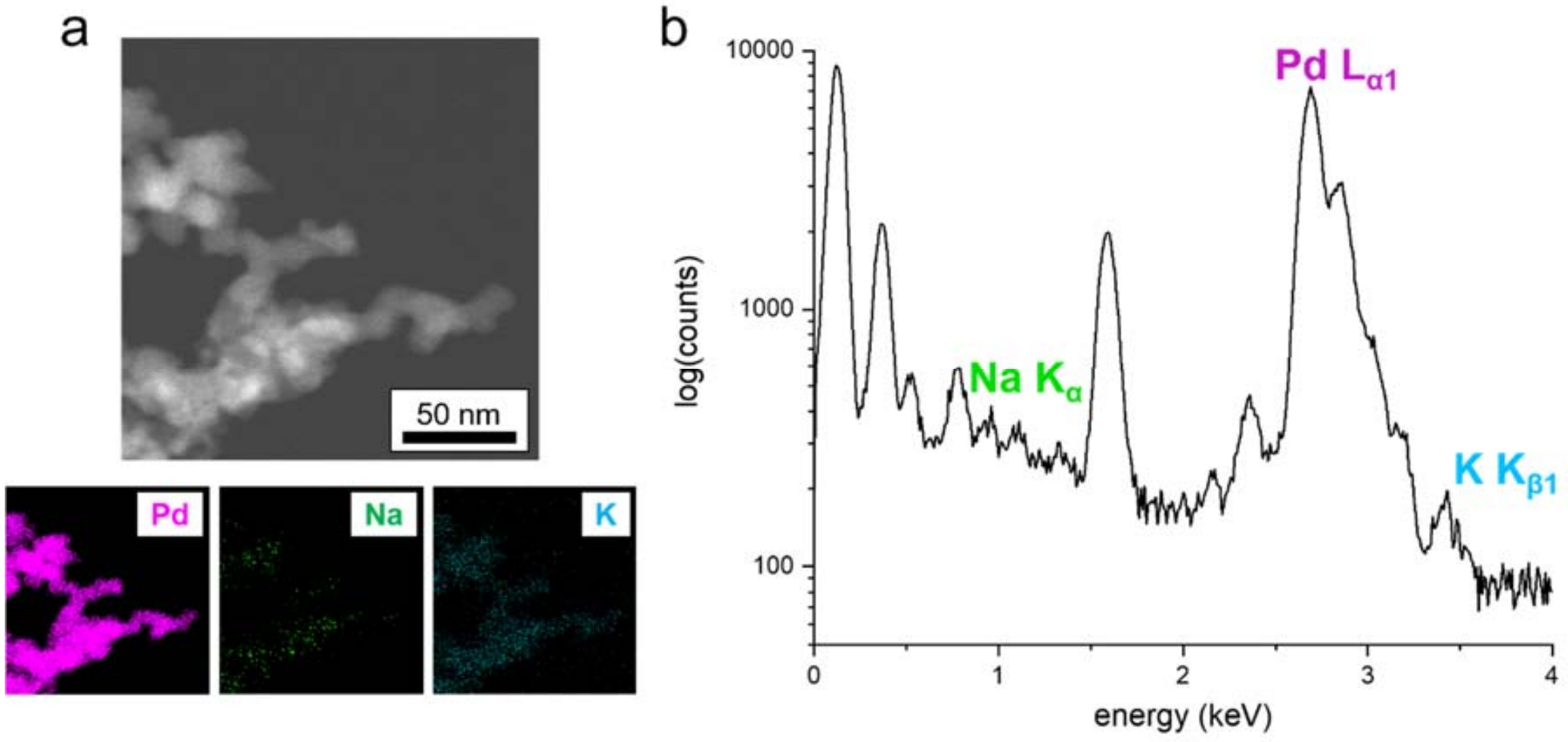

**Figure. S2.** (a) High-angle annular dark field (HAADF)- scanning transmission electron microscopy (STEM) images of the as-synthesized Pd-40 and corresponding EDS mapping of Pd (purple), Na (green), and K (cyan). (b) Overall EDS spectrum acquired from STEM.

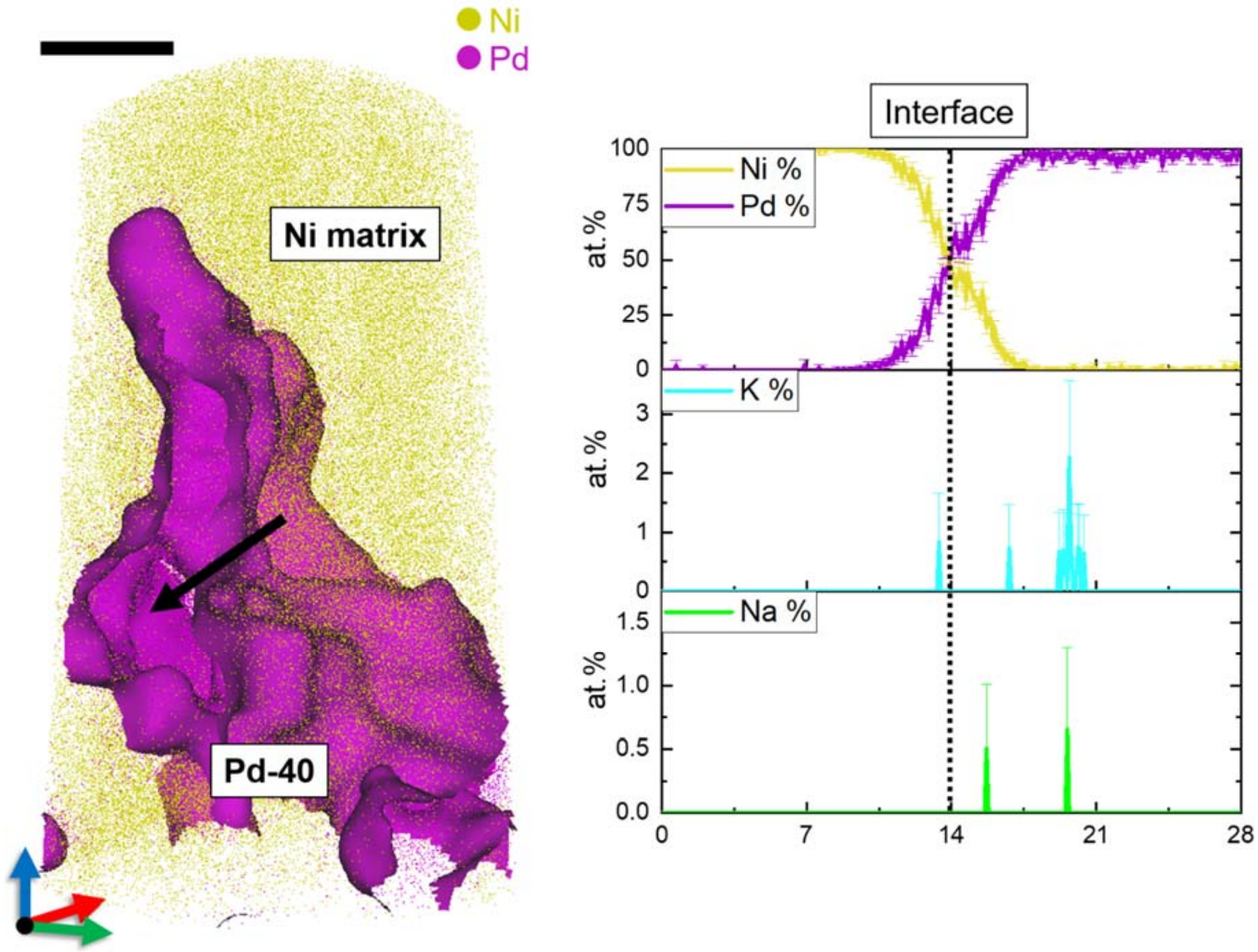

**Figure. S3.** 1D compositional profiles of the elements Ni, Pd, K, and Na contained in the Pd-40 nano-gels shown along the direction indicated by the black arrow in the 3D atom map on the left (scale bar 10 nm). The Pd/Ni interface is indicated by dotted line. A one-dimensional composition profile calculated within a cylindrical region of interest ($\Phi$5 x 28 nm$^3$, 0.1 nm bin size) positioned perpendicular to the matrix-nanomaterial interface shows that most detected impurity element, i.e. Na, and K, are present inside the Pd nano-gels not on the surface.

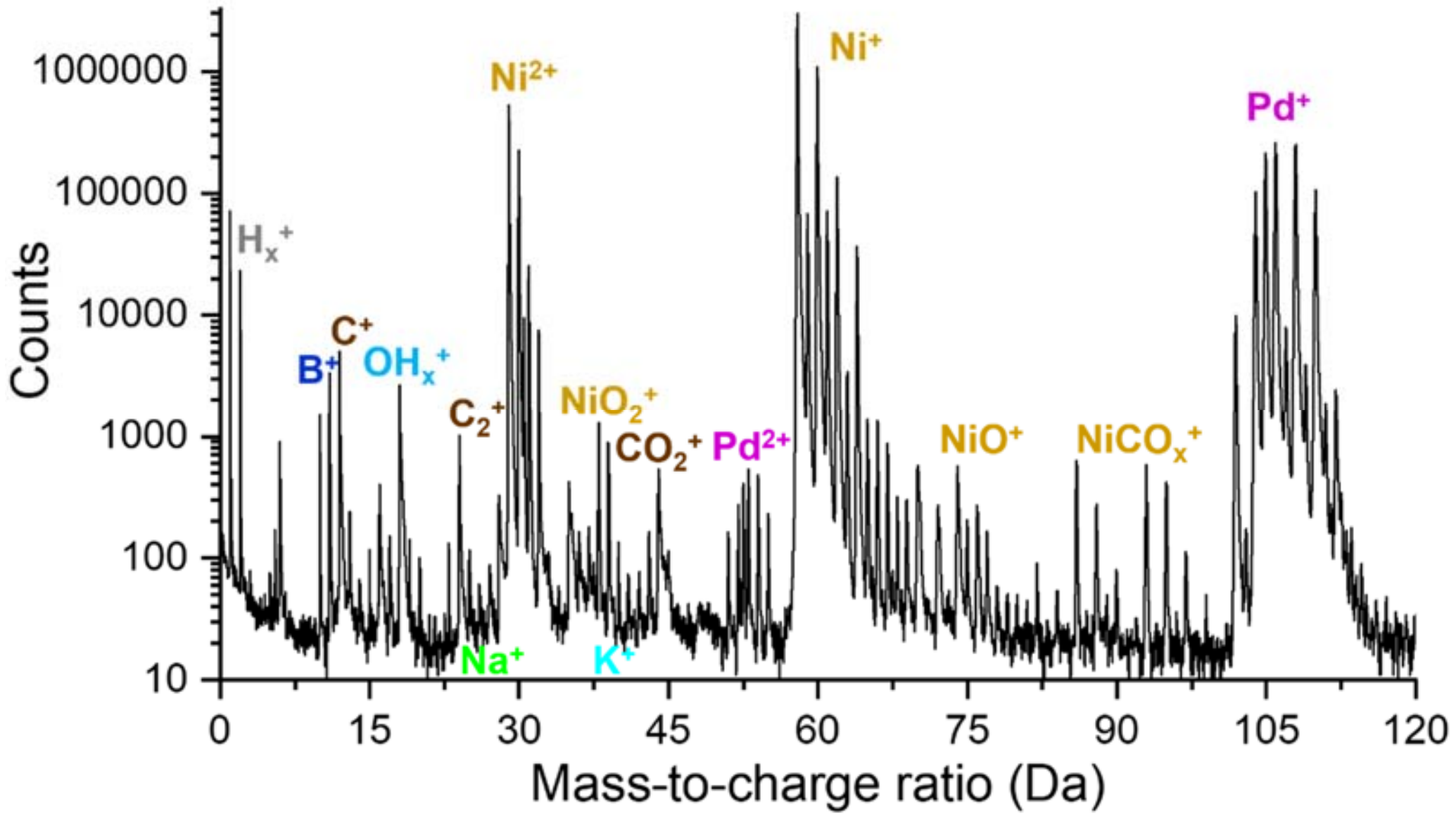

**Figure. S4.** Acquired overall mass spectrum of Pd nano-gels embedded in Ni matrix. C and O originate from H-citric acid ($C_6H_8O_7$) that was used as a buffer acid in co-electroplating sample preparation. Also B from sodium borohydride ($NaBH_4$) is detected but no segregation behavior at the grain boundaries (Figure. S11).

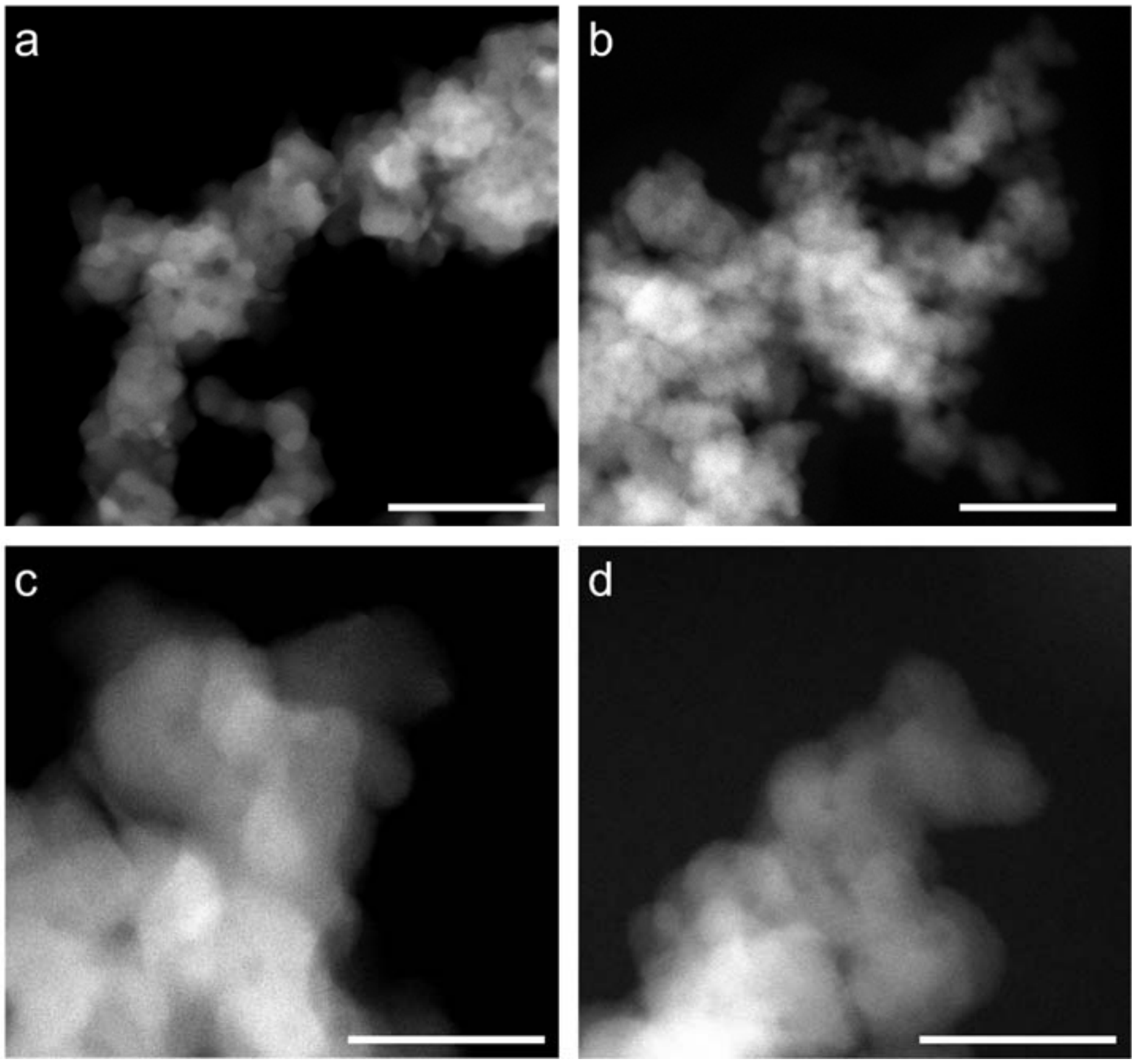

**Figure. S5.** HAADF-STEM images of (a,c) the Pd-40 and (b,d) the Pd-0.1 samples. Scale bars are 50 and 20 nm for (a,b) and (c,d), respectively.

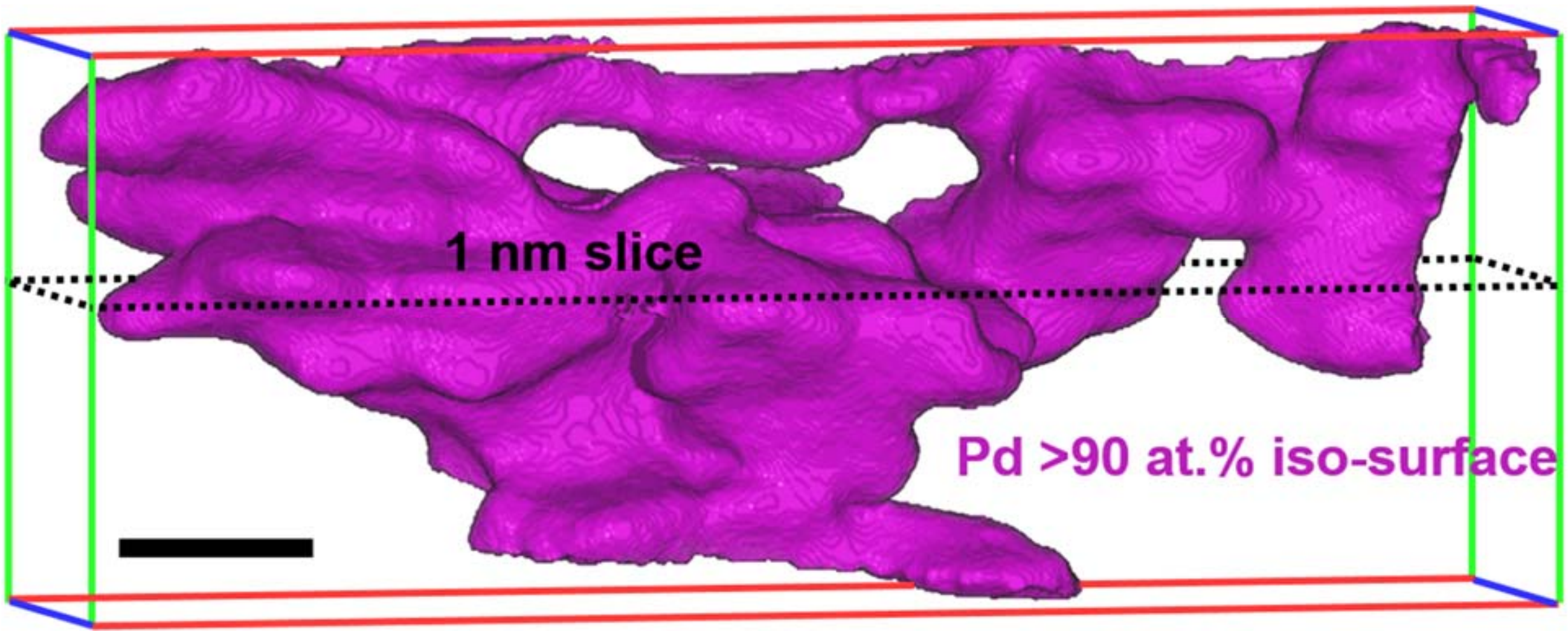

**Figure. S6.** 3D atom map of the Pd-0.1 sample. Scale bar is 20 nm. Tomogram of the 3D atom map is presented in Figure 2.

**Table S1.** Summary of the atomic composition analyses and comparison of Pd-40 and Pd-0.1 nanomaterials acquired from APT.

| Element | Pd-40 (appm) | Pd-0.1 (appm) | ratio |
|---|---|---|---|
| Na | 94 ±2 | 22 ±4 | 4.27 |
| K | 512 ±50 | 642 ±20 | 0.80 |

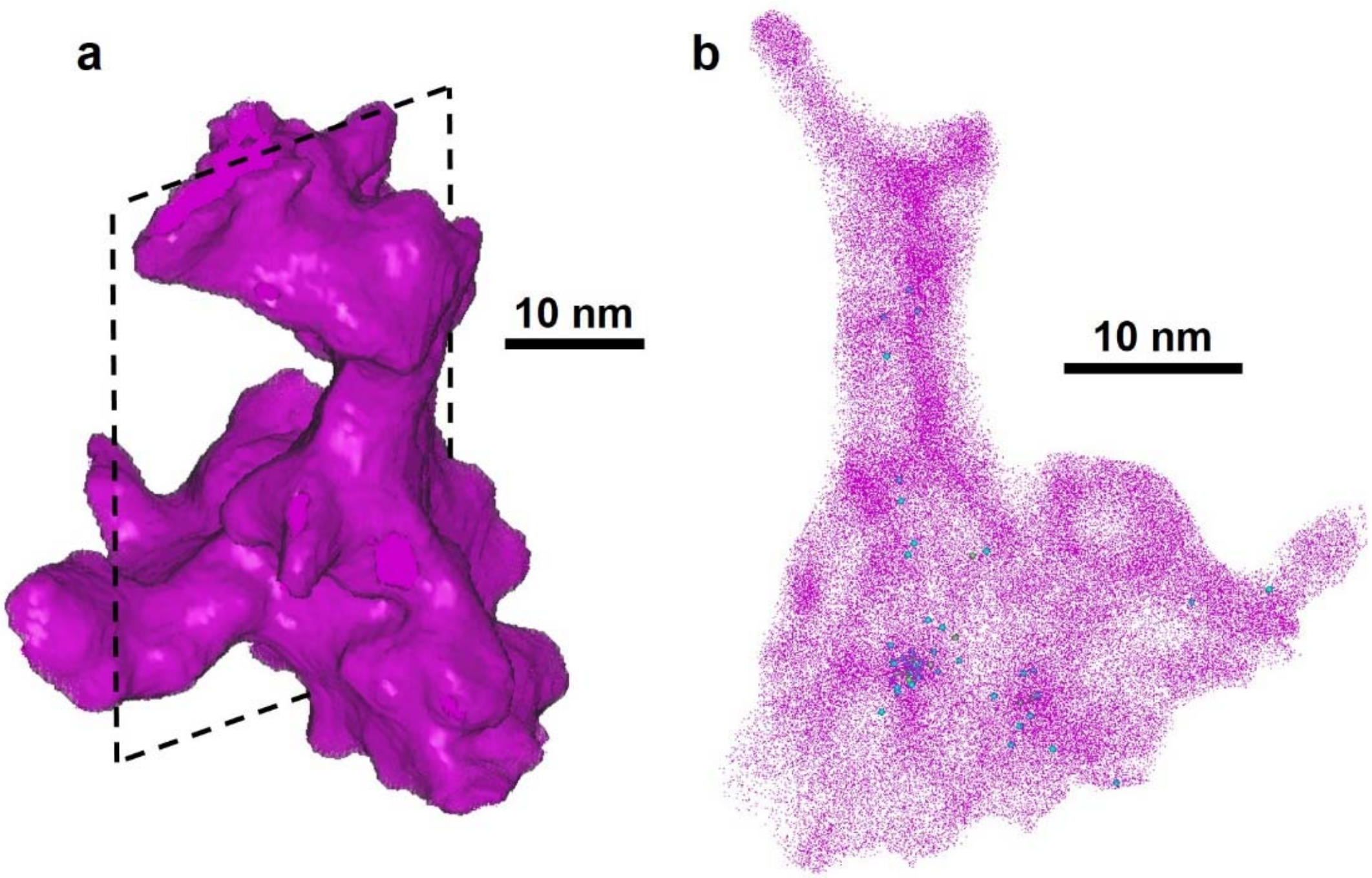

**Figure S7.** (a) 3D atom map of Pd-1 nano-gels. (b) 1-nm-thin sliced tomogram. Reconstructed Pd, Na, and K atoms are represented with purple, green, and cyan dots.

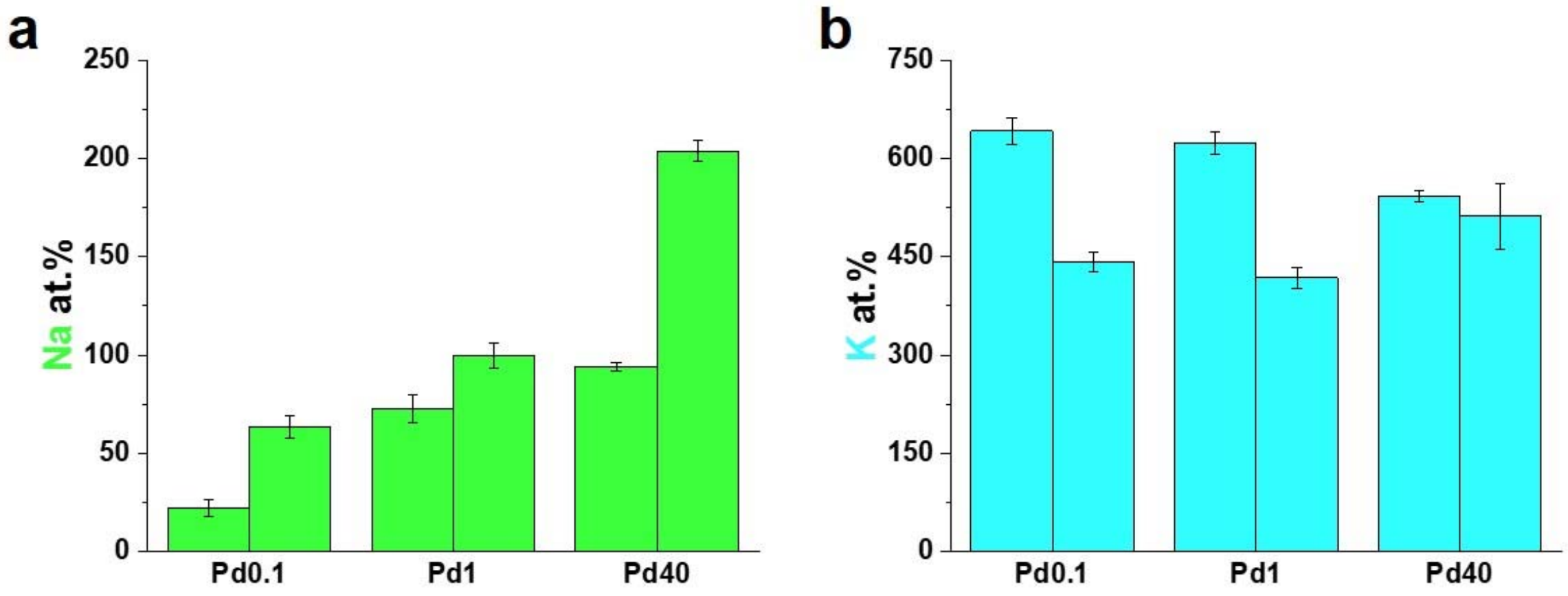

**Figure S8.** (a) Na and (b) K contents in each of the Pd gels.

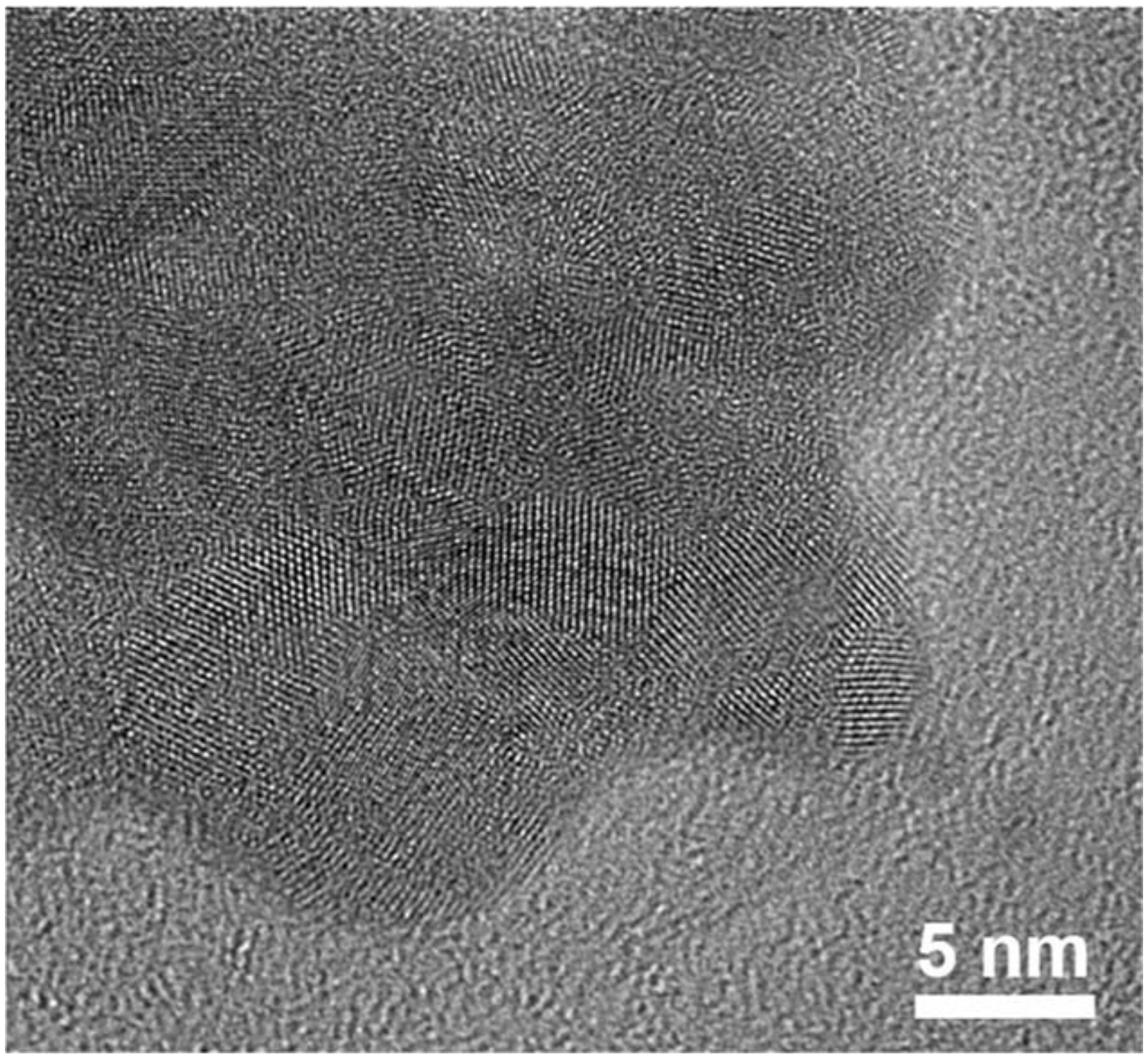

**Figure. S9.** High resolution TEM images of the Pd-0.1 sample.

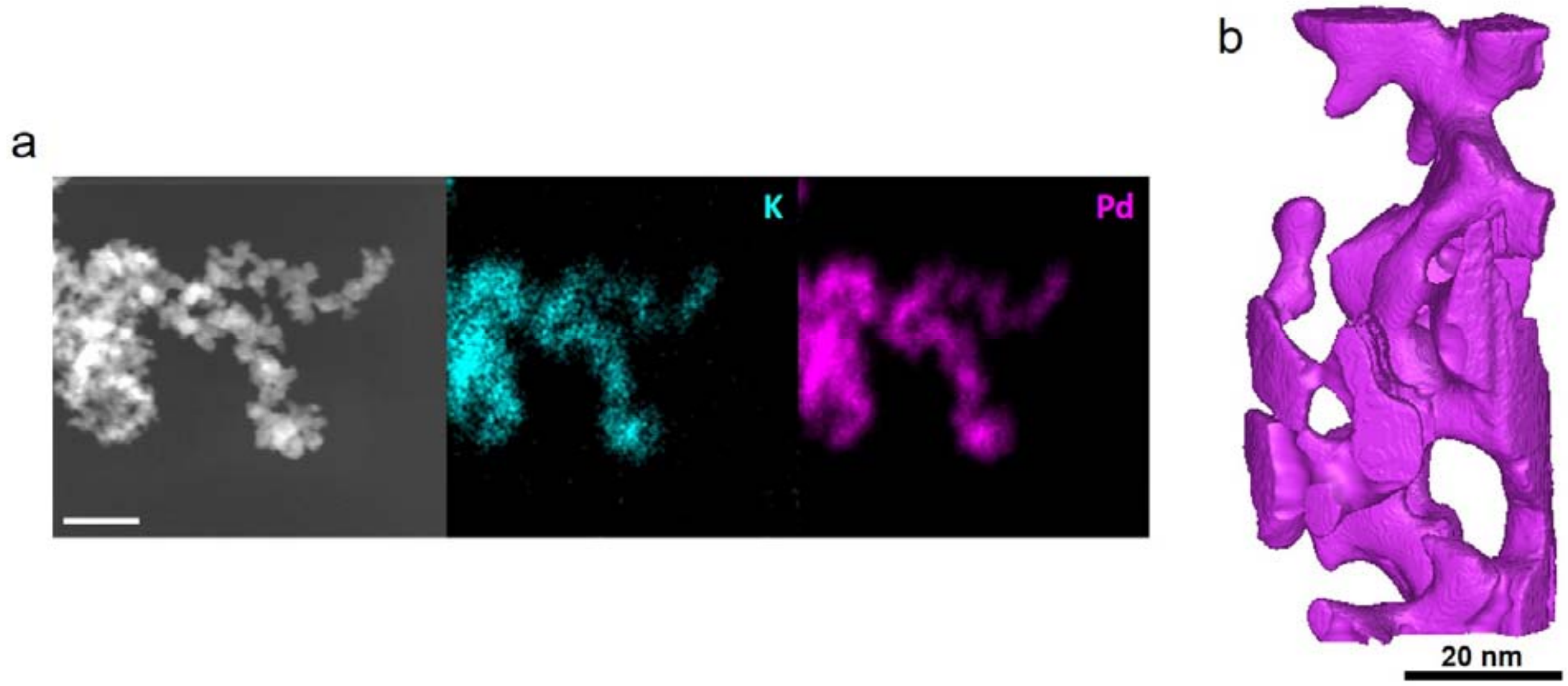

**Figure S10.** Synthesis of K-Pd-40 nano-gels to increase surface area-to-volume ratio by adding excess K during the synthesis. (a) STEM-EDS and (b) 3D atom map of the K-Pd-40 gels (white scale bar = 50 nm). KCl (>99.0%, Sigma Aldrich) was added at a 10 mole ratio level to the Pd precursor in a wet-chemical synthesis batch and followed by adding 40 mole ratio level of $NaBH_4$.

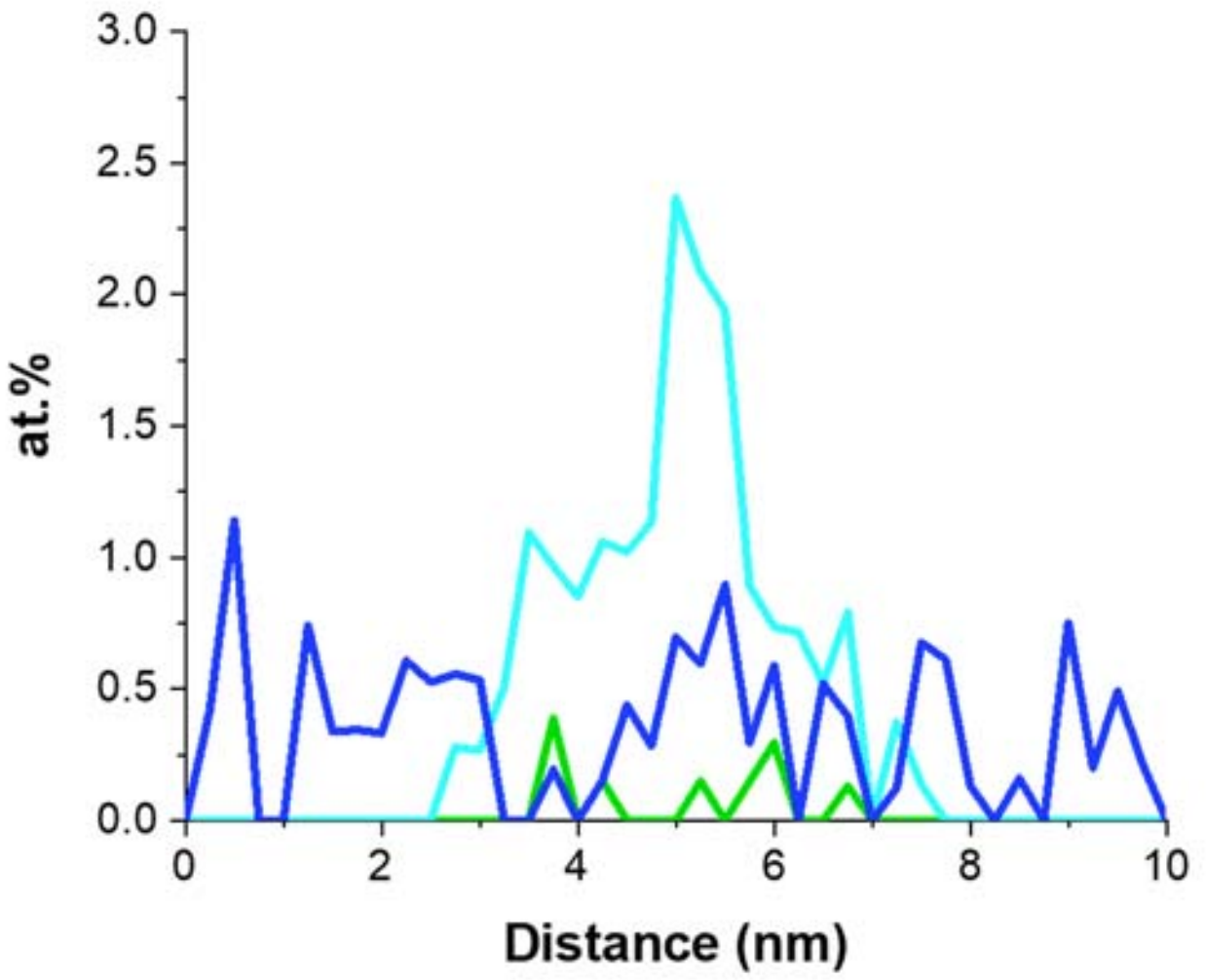

**Figure. S11.** 1D compositional profiles at the Pd-0.1 grain boundary of B (blue), Na (green), and K (cyan).

**Density Functional Theory (DFT) Calculations**

Surface adsorption calculations specific parameters

For the surface structures, a supercell containing a symmetric slab and 18 Å of vacuum is constructed. The Pd(111) slabs have 13 atomic layers (AL) and a thickness of 27.44 Å. The three outermost atomic layers are allowed to relax, while all other atoms are fixed at their bulk positions. The calculated surface energy of Pd(111) is 0.091 eV/Å$^2$, in good agreement with previous theoretical (0.082 meV/Å$^2$ [37], 0.099 meV/Å$^2$[38]) and experimental (0.125 meV/Å$^2$ [39])work.

To account for the various coverages of alkalis on Pd(111) surface, differently sized surface unit cells are employed. The coverage (Θ), defined as the ratio between the number of adsorbate atoms and the number of Pd atoms in the surface layer, has the unit of monolayer (ML). For the binding sites, we consider the top, bridge, two 3-fold hollow (i.e. FCC and HCP) and the substitutional sites. In the sub-surface region, we considered the octahedral (OSS), tetrahedral (TSS), and substitutional (SSS) sites.

The binding energy ($E_b$) of alkali adsorbate with respect to their BCC phase is calculated as

$$E_{\mathrm{b}}(c_{X^q},T) = \frac{1}{2N_X}\left[E_{\mathrm{tot}}^{X:\mathrm{Pd\,surf}} - E_{\mathrm{tot}}^{\mathrm{Pd\,surf}} - 2N_X \cdot \mu_X(c_{X^q},T)\right]$$

where $E_{\mathrm{tot}}^{X:\mathrm{Pd\,surf}}$ and $E_{\mathrm{tot}}^{\mathrm{Pd\,surf}}$ are DFT calculated total energies of the alkali-substrate system, and the Pd substrate, respectively. $N_X$ is the number of alkalis per surface cell. $\mu_X$ is the chemical potential of the alkali with respect to the BCC reference phases and accounting for the given experimental conditions, defined as

$$\mu_X(c_{X^q},T) = E_{\mathrm{tot}}^{X\,BCC} + \Delta\mu_X(c_{X^q},T)$$

where $E_{\text{tot}}^{X\,BCC}$ and $c_{X^q}$ are DFT calculated total energy of the alkali bulk (*i.e.*, BCC Na, respectively K) and the concentration of alkali ion with a charge, $q$, in solution at a given temperature, *T*. $\Delta\mu_{\text{X}}$ is the chemical potential shift for alkali ions (i.e., $K^+$ and $Na^+$) in solution with respect to the BCC reference phase calculated following ref.[3] as

$$\Delta\mu_X(c_{X^q}, T) = \Delta_f G°(X^q) - \Delta_f G(X^q)(c_{X^q}, T) + q\mu_e(c_{X^q}, T)$$

where $\Delta_f G°(X^q)$ is the formation energy of ion in the standard state with respect to the BCC bulk reference phase, which is derived by using tabulated literature data. $\Delta_f G(X^q)$ and $\mu_e$ are the formation energy of ion and the electron chemical potential at a given temperature and ion concentration. In the here discussed cases we use as input for the concentration of an alkali ion, $c_{X^q}$, the experimental values, to determine $-k_B T \ln \frac{c_{Xq}}{c_0}$ for $\Delta_f G(X^q)(c_{X^q}, T)$ where $k_B$ and $c_0$ and the Boltzmann constant and the reference concentration $c_0$ = 55.55 mol/l (considering that 1l of water contains 55.55 mol of $H_2O$ molecules). In addition, $\mu_e$ is determined utilizing the Nernst equation, which describes the change in the standard electrode potential due to changes in the concentration $q\mu_e(c_{X^q}, T) = qeU° - k_{\text{B}}T \ln \frac{c°}{c_{Xq}}$. Here $U°$ is the tabulated standard electrode potential at the reference concentration, c° = 1 mol/l. (for details, see Table S4)

Parameters specific to the grain boundary calculations

The relatively simple Σ5 (210) [100] symmetric tilt grain boundary (STGB) is selected as a representative high-angle GB in the present study. An inset in Figure. 3d shows the structure of the supercell used in all calculations. The supercell contains 40 atomic layers (4 atoms per layer, 160 atoms per cell) and represents a single cell doubled along the [100] and [012] directions. The slabs are large enough to avoid interactions between periodic images. Ionic relaxations were allowed in all calculations within a fixed shape and volume. The free surface (FS) supercell is created by removing half the number of atoms within the cell located on one side of the GB. The supercell used for the bulk and FS model calculations have the exact same dimensions as that for the GB model. In all the calculations the alkali atoms replace one, respectively two, (*i.e.* Θ is 0.25, respectively 0.50) of four equivalent Pd atoms in the GB plane. All considered GB substitutional sites are indicated in the inset of Figure. 3d. The dimension of all the models was fixed during structural optimizations allowing relaxations only along the direction perpendicular to the GB plane. All structures have been rendered using the OVITO[4] program package and all GB structures created using the software described in ref.[5].

The impact alkali atoms have on stability of the GB is analyzed by the coalescence energy given by

$$E_{\mathrm{coa}}^{X} = \left[\left(E_{\mathrm{tot}}^{X\mathrm{:Pd\ FS}} - N_{\mathrm{Pd}}^{X\mathrm{:Pd\ FS}} E_{\mathrm{tot}}^{\mathrm{Pd\ bulk}} - N_{X}^{X\mathrm{:Pd\ FS}} E_{\mathrm{tot}}^{X\ \mathrm{BCC}}\right) - \left(E_{\mathrm{tot}}^{X\mathrm{:Pd\ GB}} - N_{\mathrm{Pd}}^{X\mathrm{:Pd\ GB}} E_{\mathrm{tot}}^{\mathrm{Pd\ bulk}} - N_{X}^{X\mathrm{:Pd\ GB}} E_{\mathrm{tot}}^{X\ \mathrm{BCC}}\right)\right] / A^{\mathrm{GB}}$$

where $E_{tot}^{X\mathrm{:Pd\ GB}}$ and $E_{tot}^{X\mathrm{:Pd\ FS}}$ are the total energies of the Pd GB and the Pd FS containing $N_{\mathrm{Pd}}^{X\mathrm{:Pd\ GB}}$ and $N_{\mathrm{Pd}}^{X\mathrm{:Pd\ FS}}$ Pd atoms and $N_{X}^{X\mathrm{:Pd\ GB}}$ and $N_{X}^{X\mathrm{:Pd\ FS}}$ alkalis corresponding to 0.0 ML, 0.25 ML and 0.50 ML in the GB plane, respectively. $E_{\mathrm{tot}}^{\mathrm{Pd\ bulk}}$ and $E_{\mathrm{tot}}^{\mathrm{X\ BCC}}$ are the total energies of the Pd FCC and the alkali BCC bulk phases. $A^{\mathrm{GB}}$ is the area of the GB plane in the supercell and it is same to the area of the FS plane.

<u>Concentration analysis</u>

The thermodynamic equilibrium concentrations of alkalis in the different Pd systems (*i.e.*, on Pd surface, in Pd bulk, and at Pd GB) are calculated in areal concentration unit (*i.e.*, $cm^{-2}$). For comparison, the experimentally measured alkali contents in the MNAs are also converted to the areal concentrations for a comparison.

The upper-limit of the areal concentrations of alkalis integrated in MNAs for given experimental conditions are evaluated as

$$c^{\mathrm{upper-limit}} = \frac{\Theta^{X:\mathrm{Pd\,surf}}[E_{\mathrm{b}}(c_{X^q}, T) = 0]}{A^{\mathrm{Pd\,surf}}}$$

where $\Theta^{X:Pd\,surf}[E_{\mathrm{b}}(c_{X^q}, T) = 0]$ and $A^{\mathrm{Pd\,surf}}$ are the maximum coverage, determined as the value at which the binding energy of an alkali with respect to the ion in solution becomes zero at the given conditions, and the area of the Pd(111) $p(1\times1)$ surface unit cell, respectively.

The areal concentrations of alkalis in Pd bulk and in Pd GB plane are evaluated based on the Boltzmann distribution as follows:

$$c^{X:\alpha} = \frac{N_{\mathrm{site}} \cdot e^{\frac{-E_f^{X:\alpha}(c_{X^q}, T)}{k_{\mathrm{B}}T}}}{A^{\alpha}}$$

where $\alpha$ is the considered Pd systems [*i.e.* a Pd (4×4×4) cubic supercell with 256 Pd atoms or the Pd GB supercell with 160 Pd atoms]. $N_{\mathrm{site}}$ and $A^{\alpha}$ are the number of sites in which the alkalis can be substituted (*i.e.*, 256 for the Pd cubic bulk cell and 4 in the Pd GB plane) and the area of the corresponding system, respectively. $E_f^{X:\alpha}(c_{X^q}, T)$ is the formation energies of an alkali atom in the corresponding system at the given condition calculated as

$$E_f^{X:\alpha}(c_{X^q}, T) = E_{\mathrm{tot}}^{X:\alpha} - E_{\mathrm{tot}}^{\alpha} - [E_{\mathrm{tot}}^{X} + \Delta\mu_X(c_{X^q}, T)]$$

where $E_{\mathrm{tot}}^{\alpha}$, $E_{\mathrm{tot}}^{X:\alpha}$, and $E_{\mathrm{tot}}^{X}$ are the total energies of the pure Pd system, $\alpha$, the corresponding Pd systems with a single substitutional alkali atom, and the alkali BCC bulk reference, respectively.

$\Delta\mu_X(c_{X^q}, T)$ is the chemical potential shift for alkali ions in solution with respect to the BCC reference phase at the given conditions.

The areal concentrations of experimentally measured alkali contents in MNAs are evaluated based on the assumption that alkali atoms are present only at a single GB plane in the Pd cubic cell as follows:

$$c^{X:\mathrm{APT}} = \frac{N^{X:\mathrm{MNA}}(c_{X^q}, T)}{A^{\mathrm{cubic\ Pd}}}$$

where $N^{X:\mathrm{MNA}}(c_{X^q}, T)$ is the experimentally measured alkali contents in the MNAs at the given conditions in an appm unit and $A^{\mathrm{cubic\ Pd}}$ is an area of the assumed GB plane in the cubic Pd cell containing a milion Pd atoms. Detailed data for the concentration calculations are listed in Table S7.

**Table S2.** Calculated binding energies for Na adsorbed on Pd(111) with respect to the Na BCC bulk phase for several Na coverages Θ and different adsorption sites.

| $E_b$ (eV/Na atom) | | | | | | |
|---|---|---|---|---|---|---|
| **Θ (ML)** | **1.00** | **0.75** | **0.50** | **0.25** | **0.11** | **0.06** |
| **Top** | **0.119** | **-0.150** | **-** | **-0.888** | **-** | **-** |
| **FCC** | **-0.091** | **-0.366** | **-0.624** | **-1.047** | **-1.328** | **-1.470** |
| **HCP** | **-0.094** | **-0.340** | **-0.626** | **-1.047** | **-1.333** | **-** |
| **Bridge** | **-0.061** | **-0.360** | **-** | **-1.028** | **-** | **-** |
| **TetraI sub-surf** | **-0.028** | **-0.026** | **-** | **2.183** | **-** | **-** |
| **Octa sub-surf** | **-0.205** | **-0.154** | **0.214** | **2.513** | **-** | **-** |
| **Substitutional** | **-0.101** | **-0.421** | **-0.697** | **-0.977** | **-0.947** | **-1.050** |
| **Subst sub-surf** | **-0.215** | **-0.339** | **-0.418** | **-0.476** | **-0.325** | **-** |

**Table S3.** Calculated binding energies for K absorbed on Pd(111) with respect to the K BCC bulk phase for several K coverages Θ and different adsorption sites.

| $E_b$ (eV/K atom) | | | | | | |
|---|---|---|---|---|---|---|
| **Θ (ML)** | **1.00** | **0.75** | **0.50** | **0.25** | **0.11** | **0.06** |
| **Top** | **1.611** | **0.649** | **-0.248** | **-1.237** | **-** | **-** |
| **FCC** | **1.405** | **0.544** | **-0.336** | **-1.286** | **-1.664** | **-1.922** |
| **HCP** | **1.393** | **0.536** | **-0.335** | **-1.290** | **-1.708** | **-** |
| **Bridge** | **1.408** | **-0.250** | **-0.290** | **-1.283** | **-** | **-** |
| **TetraI sub-surf** | **1.210** | **0.681** | **-** | **1.852** | **-** | **-** |
| **Octa sub-surf** | **1.065** | **0.609** | **0.462** | **1.836** | **-** | **-** |
| **Substitutional** | **1.395** | **0.564** | **-0.155** | **-0.993** | **-1.093** | **-1.286** |
| **Subst sub-surf** | **1.058** | **0.873** | **0.921** | **1.539** | **1.842** | **-** |

**Table S4.** Tabulated cohesive energies ($E_{coh}$), standard ($T°$ = 298.15K, $p°$ = 1 bar) ionization energies ($\Delta_{IE}G°$), ion-hydration energies ($\Delta_{hyd}G°$), and standard reduction potentials ($eU°$) taken from Refs. [6,7]. The standard Gibbs free energies of formation ($\Delta_f G°$) is calculated as $\Delta_f G° = E_{coh} + \Delta_{IE}G° + \Delta_{hyd}G°$ based on Ref. [3]. The Gibbs free energy of formation ($\Delta_f G$) of ion ($Na^+$ or $K^+$) and the electron chemical potential ($\mu_e$) calculated for the chemical potential shift ($\Delta\mu_i$) of element $i$ (N or K) corresponding to the experimental conditions with respect to the BCC bulk phase are also listed.

| **Tabulated standard energies (in eV)** | | | | | | |
|---|---|---|---|---|---|---|
| **Element** | $E_{coh}$[6] | $\Delta_{IE}G°$[6] | $\Delta_{hyd}G°$[7] | $\Delta_f G°$ | $eU°$[6] | - |
| **Na** | **1.114** | **5.139** | **-3.782** | **2.471** | **-2.71** | **-** |
| **K** | **0.922** | **4.341** | **-3.057** | **2.206** | **-2.931** | **-** |
| **$\Delta_f G$, $\mu_e$ and $\Delta\mu_i$ at the given condition (in eV)** | | | | | | |
| **Ion** | $\Delta_f G$ (Pd-40) | $\mu_e$(Pd-40) | $\Delta\mu_X$ (Pd-40) | $\Delta_f G$ (Pd-0.1) | $\mu_e$ (Pd-0.1) | $\Delta\mu_X$ (Pd-0.1) |
| **Na+** | **0.128** | **-2.734** | **-0.390** | **0.282** | **-2.889** | **-0.700** |
| **K+** | **0.205** | **-3.032** | **-1.031** | **0.205** | **-3.032** | **-1.031** |

**Table S5.** Calculated binding energies for Na adsorbed on Pd(111) with respect to a $Na^+$ ion in solution at the given experimental condition for several Na coverages Θ and different adsorption sites.

| ***E*$_b$ (Pd-40) (eV/Na atom)** | | | | | | |
|---|---|---|---|---|---|---|
| **Θ (ML)** | **1.00** | **0.75** | **0.50** | **0.25** | **0.11** | **0.06** |
| **Top** | **0.509** | **0.240** | **-** | **-0.498** | **-** | **-** |
| **FCC** | **0.299** | **0.054** | **-0.234** | **-0.657** | **-0.938** | **-1.080** |
| **HCP** | **0.296** | **0.050** | **-0.236** | **-0.657** | **-0.943** | **-** |
| **Bridge** | **0.329** | **0.030** | **-** | **-0.638** | **-** | **-** |
| **TetraI sub-surf** | **0.362** | **0.364** | **-** | **2.573** | **-** | **-** |
| **Octa sub-surf** | **0.185** | **0.236** | **0.604** | **2.903** | **-** | **-** |
| **Substitutional** | **0.289** | **-0.031** | **-0.307** | **-0.587** | **-0.557** | **-0.660** |
| **Subst sub-surf** | **0.175** | **0.051** | **-0.028** | **-0.086** | **0.065** | **-** |
| ***E*$_b$ (Pd-0.1) (eV/Na atom)** | | | | | | |
| **Θ (ML)** | **1.00** | **0.75** | **0.50** | **0.25** | **0.11** | **0.06** |
| **Top** | **0.819** | **0.550** | **-** | **-0.188** | **-** | **-** |
| **FCC** | **0.609** | **0.364** | **0.076** | **-0.347** | **-0.628** | **-0.770** |
| **HCP** | **0.606** | **0.360** | **0.074** | **-0.347** | **-0.633** | **-** |
| **Bridge** | **0.639** | **0.340** | **-** | **-0.328** | **-** | **-** |
| **TetraI sub-surf** | **0.672** | **0.674** | **-** | **2.883** | **-** | **-** |
| **Octa sub-surf** | **0.495** | **0.546** | **0.914** | **3.213** | **-** | **-** |
| **Substitutional** | **0.599** | **0.279** | **0.003** | **-0.277** | **-0.247** | **-0.350** |
| **Subst sub-surf** | **0.485** | **0.361** | **0.282** | **0.224** | **0.375** | **-** |

**Table S6.** Calculated binding energies for K adsorbed on Pd(111) with respect to a $K^+$ ion in solution at the given experimental condition for several K coverages Θ and different adsorption sites.

| $E_b$ (Pd-40 = Pd-0.1) (eV/K atom) | | | | | | |
|---|---|---|---|---|---|---|
| **Θ (ML)** | **1.00** | **0.75** | **0.50** | **0.25** | **0.11** | **0.06** |
| **Top** | **2.642** | **1.680** | **0.783** | **-0.206** | **-** | **-** |
| **FCC** | **2.436** | **1.575** | **0.695** | **-0.255** | **-0.633** | **-0.891** |
| **HCP** | **2.424** | **1.567** | **0.696** | **-0.259** | **-0.677** | **-** |
| **Bridge** | **2.439** | **0.781** | **0.741** | **-0.252** | **-** | **-** |
| **TetraI sub-surf** | **2.241** | **1.712** | **-** | **2.883** | **-** | **-** |
| **Octa sub-surf** | **2.096** | **1.640** | **1.493** | **2.867** | **-** | **-** |
| **Substitutional** | **2.426** | **1.595** | **0.876** | **0.038** | **-0.062** | **-0.255** |
| **Subst sub-surf** | **2.089** | **1.904** | **1.952** | **2.570** | **2.873** | **-** |

**Table S7.** Data for the concentration analysis shown in Figure. 3 of the main text at the given systems and at the given experimental conditions.

| Pd surface | | | |
|---|---|---|---|
| **Condition** | **$\Theta^{X-\mathrm{S}}$ (ML)** | **$A^{\mathrm{Pd\ surf}}$(cm$^2$)** | **$c^{\mathrm{upper\ limit}}$** |
| **Na-Pd40** | **0.74** | **6.79×10$^{-16}$** | **1.08×10$^{15}$** |
| **Na-Pd0.1** | **0.52** | | **7.72×10$^{14}$** |
| **K-Pd40** | **0.31** | | **4.51×10$^{14}$** |
| **K-Pd0.1** | **0.31** | | **4.51×10$^{14}$** |
| **Pd bulk** | | | |
| **Condition** | **$E_f^{X:\mathrm{Pd\ bulk}}$ (eV/atom)** | **$A^{\mathrm{Pd\ bulk}}$(cm$^2$)** | **$c^{X:\mathrm{Pd\ bulk}}$** |
| **Na-Pd40** | **0.279** | **2.50×10$^{-14}$** | **2.12×10$^{11}$** |
| **Na-Pd0.1** | **0.589** | | **1.33×10$^{6}$** |
| **K-Pd40** | **2.735** | | **1.15×10$^{-30}$** |
| **K-Pd0.1** | **2.735** | | **1.15×10$^{-30}$** |
| **Pd GB** | | | |
| **Condition** | **$E_f^{X:\mathrm{Pd\ GB}}$ (eV/atom)** | **$A^{\mathrm{Pd\ GB}}$(cm$^2$)** | **$c^{X:\mathrm{Pd\ GB}}$** |
| **Na-Pd40** | **-0.194** | **6.86×10$^{-15}$** | **1.46×10$^{14}$** |
| **Na-Pd0.1** | **0.115** | | **6.73×10$^{12}$** |
| **K-Pd40** | **1.556** | | **4.21×10$^{-12}$** |
| **K-Pd0.1** | **1.556** | | **4.21×10$^{-12}$** |
| **Pd MNA** | | | |
| **Condition** | **$N^{X:\mathrm{MNA}}$ (appm)** | **$A^{\mathrm{cubic\ Pd}}$(cm$^2$)** | **$c^{X:\mathrm{APT}}$** |
| **Na-Pd40** | **94** | **6.22×10$^{-12}$** | **1.51×10$^{13}$** |
| **Na-Pd0.1** | **22** | | **3.54×10$^{12}$** |
| **K-Pd40** | **512** | | **8.23×10$^{13}$** |
| **K-Pd0.1** | **642** | | **1.03×10$^{14}$** |